\documentclass[twocolumn,showpacs,preprintnumbers,amsmath,amssymb,prl,superscriptaddress]{revtex4-1}
\usepackage{bbm}
\usepackage{mathrsfs}
\usepackage{graphicx}
\usepackage{dcolumn}
\usepackage{bm}
\usepackage{amsmath}
\usepackage{amsfonts}
\usepackage{color}
\usepackage[colorlinks=true,citecolor=blue,anchorcolor=blue]{hyperref}
\usepackage{floatrow}

\begin{document}

\title{In-Plane Magnetic-Field-Induced Quantum Anomalous Hall Plateau Transition}
\author{Jinglong Zhang}
\affiliation{State Key Laboratory of Surface Physics, Department of Physics, Fudan University, Shanghai 200433, China}
\author{Zhaochen Liu}
\affiliation{State Key Laboratory of Surface Physics, Department of Physics, Fudan University, Shanghai 200433, China}
\author{Jing Wang}
\affiliation{State Key Laboratory of Surface Physics, Department of Physics, Fudan University, Shanghai 200433, China}
\affiliation{Institute for Nanoelectronic Devices and Quantum Computing, Fudan University, Shanghai 200433, China}

\begin{abstract}
We study the critical properties of the in-plane magnetic field-induced quantum anomalous Hall (QAH) plateau transition in axion insulators. Take even septuple layer film MnBi$_2$Te$_4$ as a concrete example, we find the mirror symmetry breaking from in-plane magnetic field could induce axion insulator to QAH insulator transition, which belongs to generic integer quantum Hall plateau transition. The chiral Majorana fermion does not necessarily emerge at the QAH plateau transition in MnBi$_2$Te$_4$ due to strong exchange field, but may be quite feasible in its descendent materials MnBi$_4$Te$_7$ and MnBi$_6$Te$_{10}$.
\end{abstract}

\date{\today}


\maketitle

The magnetic topological insulators (TIs)~\cite{tokura2019} brings the opportunity to realize a large family of exotic topological phenomena~\cite{qi2008,hasan2010,qi2011,qi2009b,li2010,qi2010b,yu2010,nomura2011,wang2015b,chang2013b,mogi2017}.  One representative example is quantum anomalous Hall (QAH) effect discovered in dilute magnetic TIs at low temperature~\cite{chang2013b}. Intrinsic magnetic TIs are ideal for realizing exotic quantum states and topological phase transitions at elevated temperatures. The recent theoretical prediction~\cite{zhang2019,li2019,otrokov2018} and experimental realization of the first antiferromagnetic (AFM) TI~\cite{mong2010} in MnBi$_2$Te$_4$ has attracted intensive interest in this new class of quantum materials~\cite{gong2018,otrokov2019,lee2018,yan2019a,vidal2019,deng2019,liu2019,chen2019,yan2019b,aliev2019,hu2019,wu2019}. The even septuple layer (SL) film is predicted to be axion state with quantized topological magnetoelectric effect (TME)~\cite{zhang2019}. The axion state has been experimentally observed in 6 SL MnBi$_2$Te$_4$ with vanishing Hall resistance $\rho_{xy}$ and large longitudinal resistance $\rho_{xx}$, where an out-of-plane magnetic field could drive it into a state with $\rho_{xy}\rightarrow\pm h/e^2$ and $\rho_{xx}\rightarrow0$~\cite{deng2019,liu2019}. The plateau transition is of particular interest, which may provide a platform for chiral Majorana fermion mode (CMFM) based quantum computing by proximity coupling to $s$-wave superconductor~\cite{qi2010b,wang2015c,lian2018,lian2017}. However, the large out-of-plane field will destroy the superconductivity. While the in-plane critical field of 2D superconductors are found to be much larger than that of out-of-plane~\cite{poole2014,lu2015,xi2016}. It is natural to ask whether in-plane field could induce QAH plateau transition. If it does, what are the critical properties of quantum phase transition? Furthermore, whether it could be used for realizing CMFM. In this paper, we address these issues by studying even SL MnBi$_2$Te$_4$ as a concrete example, which is generic for magnetic TIs.

\begin{figure}[t]
\begin{center}
\includegraphics[width=3.1in,clip=true]{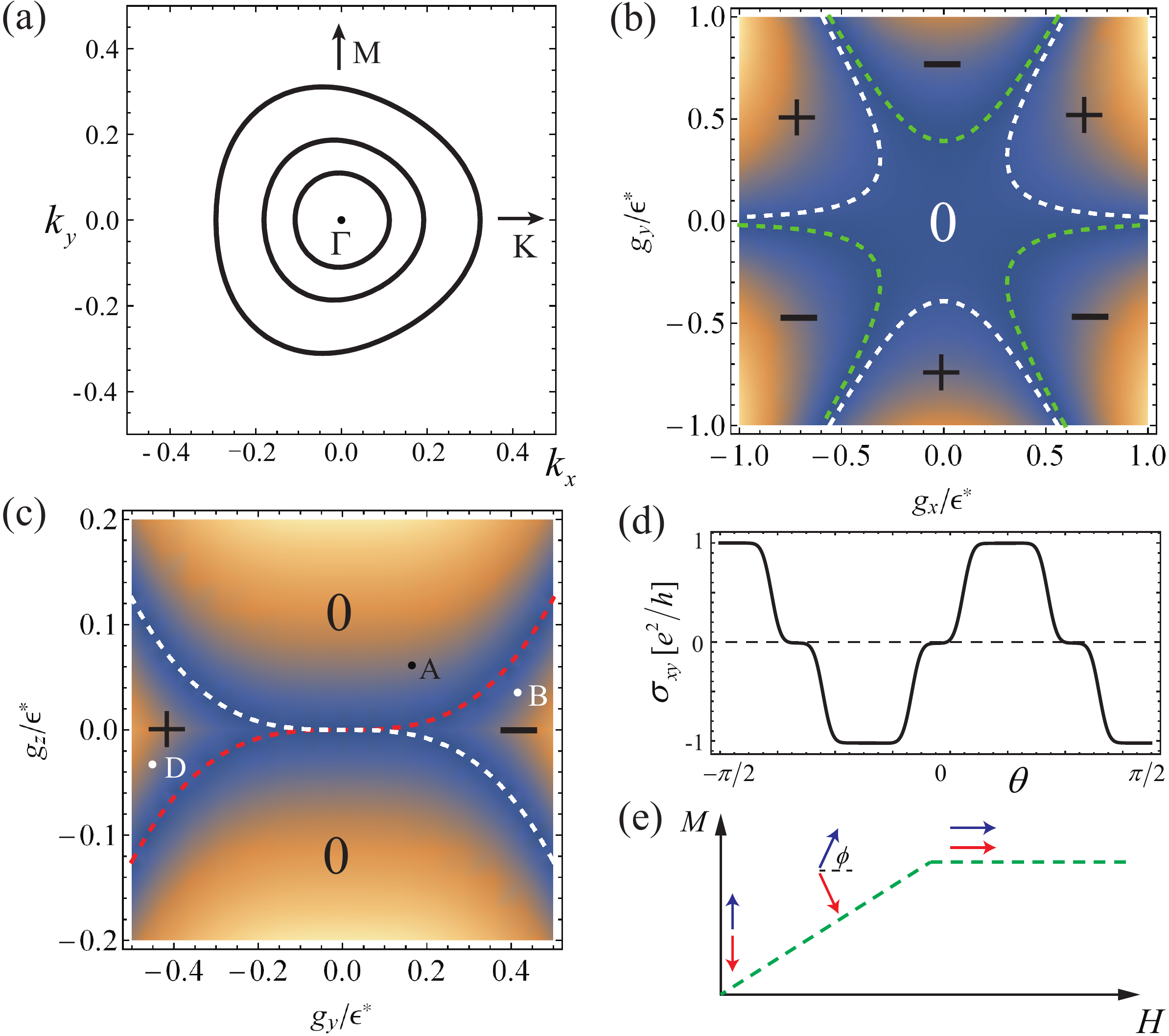}
\end{center}
\caption{(a) Constant energy contour of $H(\vec{k})$. From inner to outer, the energy is $0.2\epsilon^*$, $0.25\epsilon^*$, $0.35\epsilon^*$. $k_x$ and $k_y$ are in unit of $\sqrt{v/\lambda}$. $\epsilon^*=0.3$~eV, $g_z=0.05$~eV~\cite{vidal2019}. (b) Phase diagram in $g_x-g_y$ plane with $g_z/\epsilon^*=0.06$. $\pm$ denotes $C=\pm1$. (c) Phase diagram in $g_y-g_z$ plane with $g_x=0$. (d) Sketch of Hall conductance versus the angle $\theta$ of in-plane magnetic field respect to $x$ axis when $\mu$=0. (e) Diagrammatic representation of AFM manipulation by in-plane magnetic field.}
\label{fig1}
\end{figure}

\emph{Model.} The (111) Dirac surface state of MnBi$_2$Te$_4$ is gapped due to time-reversal $\Theta$ breaking. The noncircular Fermi surface of surface states observed in ARPES~\cite{vidal2019} is from the-threefold warping term~\cite{fu2009}, where the surface model is $H(\vec{k})=v(k_y\sigma_x-k_x\sigma_y)+(\lambda/2)(k_+^3+k_-^3)\sigma_z+g_z\sigma_z$. Here $v$ is the Dirac velocity, $k_\pm=k_x\pm ik_y$ with $x$ axis along $\Gamma K$, $\lambda$ is the warping parameter, $\sigma_i$ are Pauli matrices acting on spin space,  $g_z=J_z\langle S_z\rangle$ is the surface Zeeman term due to exchange field along $z$ axis introduced by surface ferromagnetic ordering, $\langle S_z\rangle$ is the mean field expectation value of surface local spin along $z$ axis, $J_z<0$ is the effective exchange parameter between local moment and band electron. For simplicity, the particle-hole asymmetry is neglected. Defining the characteristic energy $\epsilon^*=v\sqrt{v/\lambda}$ and wavevector $\sqrt{v/\lambda}$, we plot in Fig.~\ref{fig1}(a) a set of constant energy contours within the bulk gap, consistent with the first-principles calculations~\cite{zhang2019}.

Now we turn to zero Hall plateau state in even SL film. The low energy physics is described by the massive Dirac surface states only, where the intrinsic N\'{e}el-type ordering introduces opposite Zeeman term on two surfaces. The generic form of the effective Hamiltonian is
\begin{eqnarray}\label{single}
\mathcal{H}(\vec{k})  &=& v(k_y\sigma_x-k_x\sigma_y)\tau_z+\frac{\lambda}{2}(k_+^3+k_-^3)\sigma_z\tau_z
\nonumber
\\
&&+g_z\sigma_z\tau_z+g_x\sigma_x+g_y\sigma_y,
\end{eqnarray}
with the basis of $|t\uparrow\rangle$, $|t\downarrow\rangle$, $|b\uparrow\rangle$, and $|b\downarrow\rangle$, where $t$ and $b$ denote the top and bottom surface states and $\uparrow$ and $\downarrow$ represent spin up and down states, respectively. The Pauli matrices $\tau_i$ act on layer. We neglect the hybridization between two surfaces, which is also negligible when film exceeds 4 SL. $\vec{g}_{\parallel}\equiv(g_x, g_y)=J_{\parallel}(\langle S_x\rangle,\langle S_y\rangle)$ is in-plane Zeeman-type exchange field, which can originate from exchange field due to magnetization of Mn induced by in-plane magnetic field, or the direct Zeeman coupling between band electron and magnetic field. In the absence of field, the system is AFM with N\'{e}el order along $z$ axis. When the in-plane field is applied [Fig.~\ref{fig1}(e)], the magnetic moments on two sublattices cant and a net magnetization gradually builds up proportional to the field as described by the Stoner-Wohlfarth model~\cite{baltz2018}. Interestingly, $g_i$ ($i=x,y,z$) is continuously tunable. When $\lambda=\vec{g}_{\parallel}=0$, the system is zero Hall plateau state with quantized topological axion response~\cite{wang2015b,zhang2019}, which is also called axion insulator~\cite{mogi2017}. Here we focus on $\lambda, \vec{g}_{\parallel}\neq0$, which lead to in-plane field-induced QAH plateau transition.

\emph{Phase diagram.} A general symmetry analysis on the Hall conductance will help us to understand intuitively the in-plane field-induced QAH state. $\Theta$ breaking is necessary for
nonzero $\sigma_{xy}$, which always exist in this system. Besides, $\mathcal{I}\Theta$ symmetry constrains $\sigma_{xy}=0$, which corresponds to $\vec{g}_{\parallel}=0$. Here $\mathcal{I}$ is inversion operator. A finite $\vec{g}_{\parallel}$ leads to $\mathcal{I}\Theta$ breaking. Furthermore, the mirror symmetry in 2D also leads to $\sigma_{xy}=0$~\cite{fang2012,liu2013}. The MnBi$_2$Te$_4$ film has three mirror-symmetric $\Gamma M$ direction if $g_z=0$, thus, the pseudovector $\vec{g}_{\parallel}$ should not be perpendicular to $\Gamma M$ (namely, not to parallel to $\Gamma K$) for nonzero $\sigma_{xy}$. Therefore, as $g_i$ is continuously tuned by in-plane field, the QAH plateau transition is expected.

The Hamiltonian in Eq.~(\ref{single}) is classified by the Chern number $C$. Since the topological invariants cannot change without closing the bulk gap, the phase diagram can be determined by first finding the phase boundaries as gapless regions in parameter spaces, and then calculate $C$ of the gapped phases. The two surfaces in $\mathcal{H}(\vec{k})$ decouple with the band dispersion $E_{t/b}=\pm\sqrt{(g_z+\lambda(k^3_x-3k_xk_y^2))^2+(g_y\mp vk_x)^2+(g_x\pm vk_y)^2}$, with the gap closing point at $(k_x,k_y)=\pm(g_y/v,-g_x/v)$ and $g_z=-\lambda(k^3_x-3k_xk_y^2)$. This leads to the phase diagram shown in Fig.~\ref{fig1}(c). Point $A$ is adiabatically connected to $g_z\neq0$ and $\vec{g}_{\parallel}=0$ limit with $C=0$. While point $B$ is adiabatically connected to $g_z=0$ and $g_y\neq0$ with $C=-1$~\cite{liu2013}. This can be understood by adding a small perturbation $g_z'\sigma_z$ into Eq.~(\ref{single}), and the system is further adiabatically connected to $g_z'<0$ and $g_y=0$, where $C=g_z'/|g_z'|=-1$~\cite{wang2014a}. Similar analysis can be applied to point $D$. The Chern number of all gapped regimes is further determined from the $C_{3z}$ rotational symmetry, as shown in Fig.~\ref{fig1}(b).

Explicitly, the phase diagram can be understood from phase transition of surface Dirac model. In Fig.~\ref{fig1}(c), the white (red) line corresponds to the phase transition from top (bottom) surfaces. The gap of the top surface occurs at $(k_{t}^x,k_{t}^y)=(g_y/v,-g_x/v)$, where the effective model is re-written as $H_t=m_t'\sigma_z+k_{t}^{y\prime}\sigma_x-k_{t}^{x\prime}\sigma_y$, with $k_{t}^{x\prime}=k_t^x-g_y/v$,  $k_{t}^{y\prime}=k_t^y+g_x/v$ and $m_t'=g_z+g_y^3-3g_x^2g_y$. Such continuous Dirac model has half-quantized Hall conductance due to meron-type configuration in $(k_x',k_y')$ space~\cite{redlich1984,qi2008}. Namely
\begin{equation}\label{Hall}
\sigma_{xy}^t\equiv C_t\frac{e^2}{h}=-\frac{\text{sgn}(m_t')}{2}\frac{e^2}{h}.
\end{equation}
While for bottom surface, $H_b=m_b'\sigma_z-k_b^{y\prime}\sigma_x+k_b^{x\prime}\sigma_y$ with $k_{b}^{x\prime}=k_b^x+g_y/v$,  $k_{b}^{y\prime}=k_b^y-g_x/v$  and $m_b'=g_z-g_y^3+3g_x^2g_y$, thus $C_{b}=-\text{sgn}(m_b')/2$. Therefore, the total Chern number $C=C_t+C_b$.

\begin{figure}[b]
\begin{center}
\includegraphics[width=2.9in,clip=true]{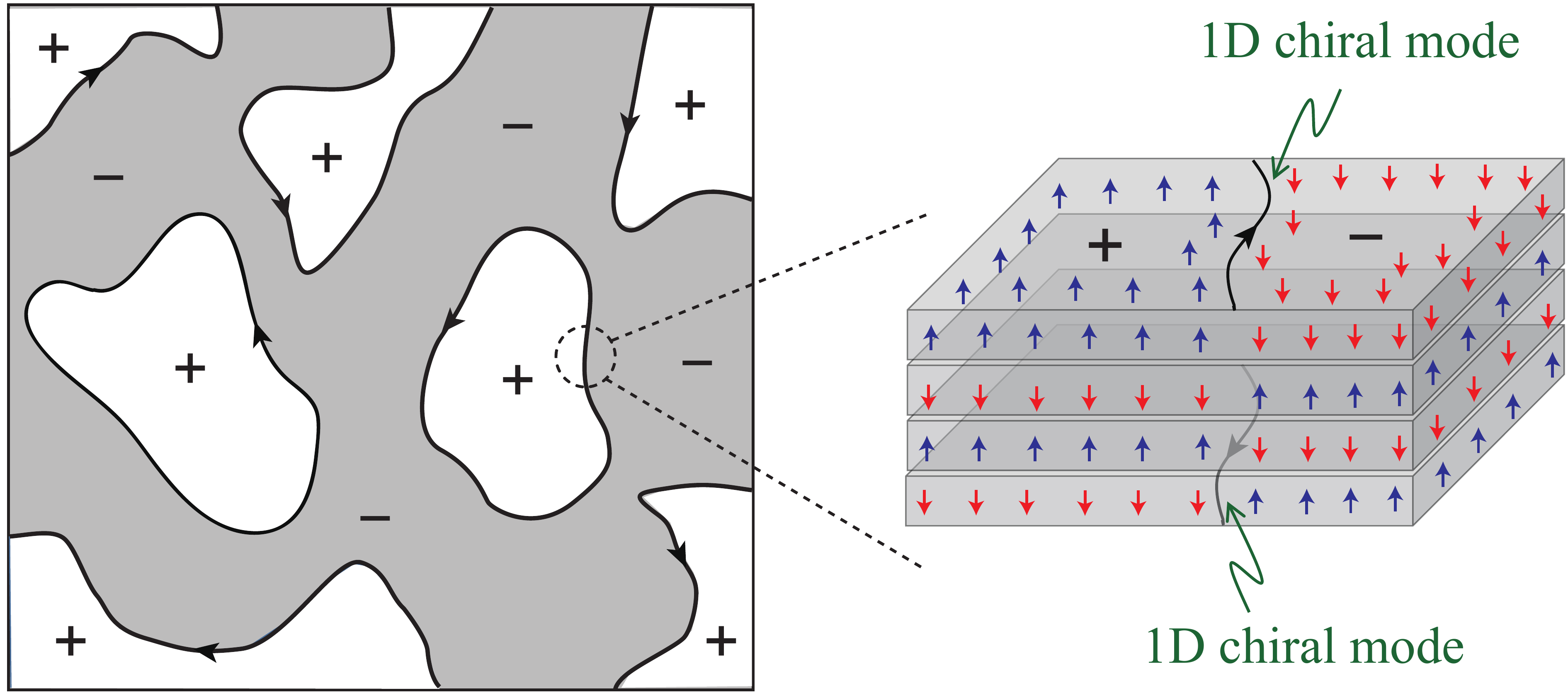}
\end{center}
\caption{AFM domains generically exist in even SL MnBi$_2$Te$_4$ film. There exist 1D chiral states (indicated by arrow lines) along AFM domain walls on $t,b$ surfaces. On left panel only the top surface is shown. The symbols $+$ (white region) and $-$ (gray region) denote the upward ($\uparrow\downarrow\uparrow\downarrow$) and downward ($\downarrow\uparrow\downarrow\uparrow$) magnetic domains. The bottom surface is shown on right panel, where the chirality of 1D mode reverses.}
\label{fig2}
\end{figure}

\emph{Domain.} The above analysis on Chern number of surface Dirac model gives us a clear picture of phase diagram in a uniform AFM system. AFM domains ($\uparrow\downarrow$ and $\downarrow\uparrow$) are constructed because they are degenerate energetically. If N\'{e}el ordering starts at one point and develops to the whole crystal, there is no stray field and will be only one domain. Ordinarily, however, that is not the case. The crystalline imperfection is the common reason for AFM domain formation. Even in the perfect crystal, the lowering of free energy that accompanies an increase in entropy can lead to an equilibrium multidomain structure as shown in Fig.~\ref{fig2}.

The opposite AFM domains have opposite TME, thus even SL MnBi$_2$Te$_4$ has much reduced even \emph{vanishing} axion response due to multidomains. Fortunately, for the magnetoelectric crystals here, there are several ways to differentiate one AFM domain from the other. One simple way is to apply the electric field on MnBi$_2$Te$_4$, cooled below N\'{e}el temperature $T_{N}$ without magnetic field, one can expect that the directions of induced orbital magnetic moments are different from domain to domain. The AFM domain structure can be visualized by observing polarity of the induced magnetic moment by Kerr technique. Another way is to use the second harmonic generation to measure N\'eel ordering~\cite{fiebig2005}. There exist 1D gapless chiral mode at AFM domain walls on both Dirac surface states as shown in Fig.~\ref{fig2}, where the chiralities are opposite. This offer another way to differentiate the domains by imaging the conducting 1D chiral modes, through scanning tunnelling microscope or microwave impedance microscopy~\cite{lai2010}.

The intrinsic quantized TME in even SL MnBi$_2$Te$_4$ can be measured when the AFM domains are eliminated. In this case, the system is an axion insulator instead of a normal insulator. This can be achieved by the magnetoelectric field cooling with magnetic and electric fields applied simultaneously~\cite{borisov2005}, which favors a distinct AFM single domain.

\emph{Plateau transition.} From Eq.~(\ref{Hall}), by varying $m_t'$ from some negative value to a positive value, we see a jump from $1/2$ to $-1/2$ in $\sigma_{xy}/(e^2/h)$. While the Dirac mass of bottom surface does not change sign, implies the Hall plateau transition from $0$ to $-1$ in these units. Similarly, the bottom surface is responsible for $1$ to $0$ transition when $m_b'$ changes sign. They are nondegenerate as long as $g_z\neq0$. By applying in-plane magnetic field, let's say along $\Gamma M$ direction, the QAH plateau transition happen at opposite fields with $m_t'=0$ and $m_b'=0$, respectively. The quenched disorder will generate spatially random perturbations to Eq.~(\ref{single}). There generically exist three types of randomness,
\begin{eqnarray}
\mathcal{H}^j_{g} &=& g^j_z(x,y)\sigma_z+g^j_x(x,y)\sigma_x+g^j_y(x,y)\sigma_y,
\nonumber
\\
\mathcal{H}^j_{A} &=& A^j_x(x,y)\sigma_y-A^j_y(x,y)\sigma_x,
\\
\mathcal{H}^j_{V} &=& V^j(x,y).
\nonumber
\end{eqnarray}
where $j=t,b$ simply means the two surfaces may feel different randomness. $\vec{A}^j\equiv(A^j_x, A^j_y)$, $\vec{g}^j\equiv(g^j_x, g^j_y, g^j_z)$, and $V^j$ are nonuniform and random in space but constant in time. $\mathcal{H}_g$ corresponds to random exchange field induced by local spin in magnetic domains. $\mathcal{H}_A$ is a random vector potential, which comes from the gauge coupling $(\vec{k}\rightarrow\vec{k}-\vec{A})$ with random stray magnetic field in the system. $\mathcal{H}_V$ is the random scalar potential induced by impurities in the material. To be concrete, at $m_{t,b}'=0$, we assume that all three random potentials are symmetrically distributed about zero mean. We also assume the interaction between the electrons can be neglected.

\begin{figure}[t]
\begin{center}
\includegraphics[width=2.2in,clip=true]{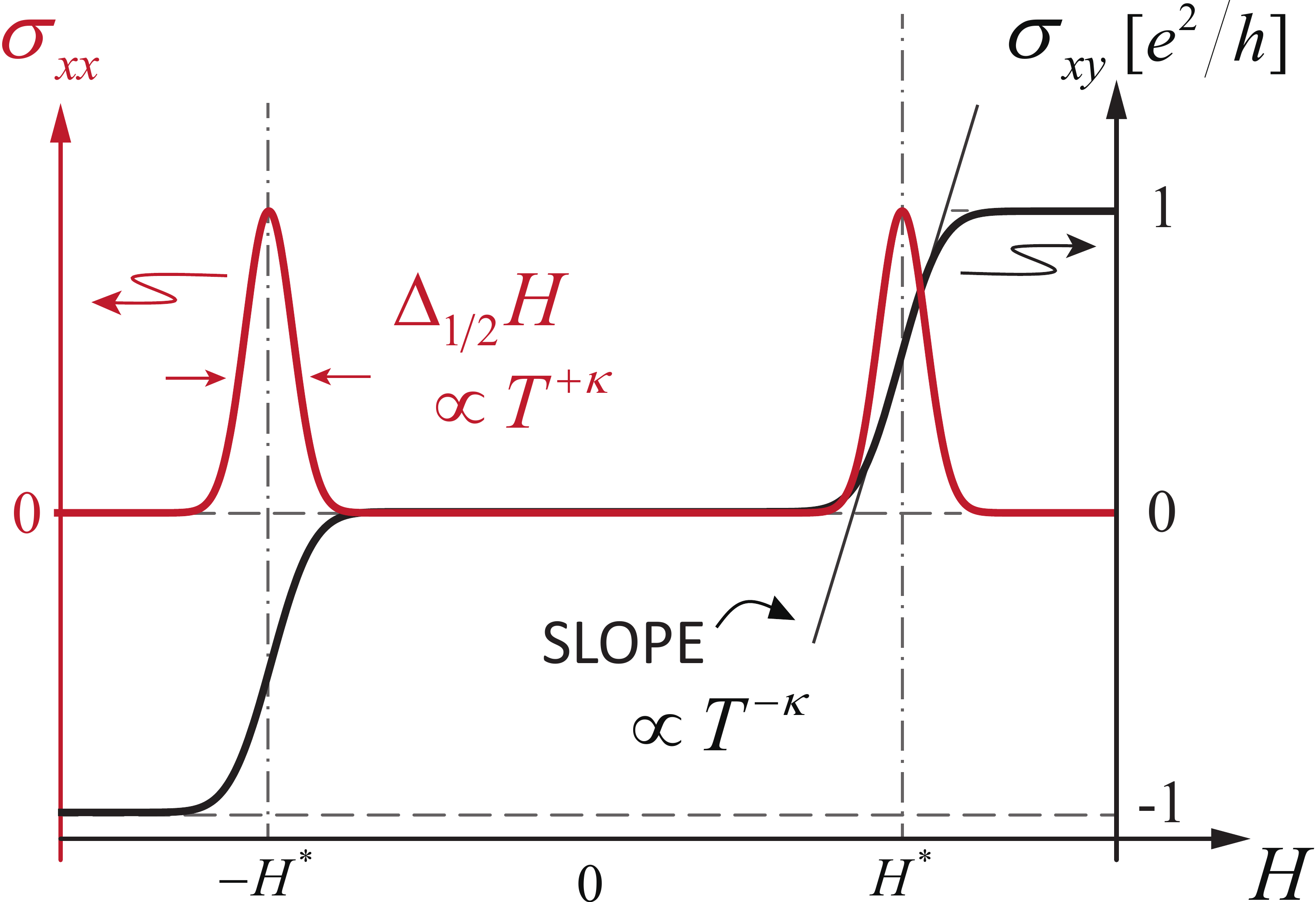}
\end{center}
\caption{Sketch of in-plane magnetic field dependence of $\sigma_{xy}$ and $\sigma_{xx}$.}
\label{fig3}
\end{figure}

If the system has only single AFM domain, then the in-plane field-induced QAH plateau transition here is exactly the \emph{doubled} version of random Dirac model for the integer QHE transition~\cite{fisher1985,ludwig1994}. The fixed point of random Dirac model is first conjectured to be a generic integer QHE fixed point~\cite{ludwig1994,kivelson1992,huckestein1995,sondhi1997,kramer2005}, and later confirmed by exact mapping to the network model~\cite{chalker1988,ho1996}. The mapping between the doubled Dirac model and network model has been studied in Ref.~\cite{wang2014a}. Therefore, the critical exponent obtained for the latter~\cite{slevin2009} can be used for the former. The AFM multidomains introduce extra complications. There are three distinct cases. (i) $\uparrow$ and $\downarrow$ domains dominate on the top and bottom surface, respectively, then QAH plateau transition is the same to the single domain case. Namely, the in-plane field-induced $1/2$ to $-1/2$ transition on the top surface, while the bottom surface remains to be $-1/2$. (ii) $\uparrow$ domains dominate on both of two surfaces. The field induces $1/2$ to $-1/2$ transition on both surfaces. However, due to different random perturbations, the transitions on two surface are generically non-degenerate. The system will experience discrete $1$ to $0$, then to $-1$ transition. (iii) $\uparrow$ and $\downarrow$ domains are the same, and two surfaces are at critical point. Then the system is no longer an insulator, but an critical metal with 1D \emph{helical} modes percolating. In this state $\sigma_{xy}=0$ due to averaged $\mathcal{I}\Theta$, but $\sigma_{xx}$ is finite. A small in-plane field will drive the system into case (i). Therefore, AFM multidomains will not affect the critical behavior of in-plane field-induced QAH plateau transitions.

The critical phenomena in above QAH plateau transition implies universal finite-size scaling behaviour in the conductance and resistance matrices. More specifically, Fig.~\ref{fig3} shows the in-plane magnetic field dependence of $\sigma_{xy}$ and $\sigma_{xx}$. There exist two critical points at $\pm H^*$ at which the localization length $\xi\propto|H-H^*|^{-\nu}$ diverges. The critical exponent $\nu\approx2.4$~\cite{li2009,slevin2009}, $H^*$ is the critical external field of the plateau transition. The single parameter scaling~\cite{pruisken1988} suggest the maximum slope in $\sigma_{xy}$ diverges as a power law in temperature as $(\partial\sigma_{xy}/\partial H)_{\text{max}}\propto T^{-\kappa}$. While the half-width of $\sigma_{xx}$ peak vanishes like $\Delta_{1/2}H\propto T^{\kappa}$~\cite{shklovskii1993}. Here $\kappa=p/2\nu$, and $p$ is determined from phase coherence length $L_{\text{in}}\propto T^{-p/2}$~\cite{thouless1977}. The statement for $\sigma_{\alpha\beta}$ can be directly translated into resistance $\rho_{\alpha\beta}$ through $\sigma_{\alpha\beta}=\rho_{\alpha\beta}/(\rho_{xx}^2+\rho_{xy}^2)$. Still, one can observe two Hall resistance plateau transitions at $\pm H^*$, with $(\partial\rho_{xy}/\partial H)_{\text{max}}\propto T^{-\kappa}$. However, $\rho_{xx}$ will become a \emph{single} peak due to insulating state at zero Hall plateau, where around the critical field, $\rho_{xx}=f[(H-H^*)T^{-\kappa}]$ with $f$ a regular function. Moreover, by rotating the in-plane field, $\sigma_{xy}/(e^2/h)$ will switch between $1$, $0$, $-1$, depending on the angle between in-plane field and crytalline orientation, and the above scaling behaviors also applies.

\emph{Chiral TSC.} The chiral topological superconductor (TSC) with odd $\mathcal{N}$ of CMFM was proposed to generically emerge at QAH plateau transition in proximity to $s$-wave superconductor~\cite{qi2010b}. This motivates us to study the phase diagram of the above system when proximity coupled to superconductor. The Bogoliubov-de Gennes (BdG) Hamiltonian is $H_{\text{BdG}}=(1/2)\sum_{\vec{k}}\Psi^\dag_{\vec{k}}\mathcal{H}_{\text{BdG}}\Psi_{\vec{k}}$, with $\Psi_{\vec{k}}=(\psi_{\vec{k}}^T,\psi^\dag_{-\vec{k}})$, $\psi_{\vec{k}}=(c^t_{\vec{k}\uparrow}, c^t_{\vec{k}\downarrow}, c^b_{\vec{k}\uparrow},c^b_{\vec{k}\downarrow})$, and
\begin{eqnarray}
\mathcal{H}_{\text{BdG}}(\mathbf{k}) &=&
\begin{pmatrix}
\mathcal{H}(\vec{k})-\mu & \Delta(\vec{k})
\\
\Delta^\dag(\vec{k}) & -\mathcal{H}^*(-\vec{k})+\mu
\end{pmatrix},
\nonumber
\\
\Delta(\vec{k}) &=&
\begin{pmatrix}
i\Delta_t\sigma_y & 0
\\
0 & i\Delta_b\sigma_y
\end{pmatrix}.
\end{eqnarray}
Here $\mu$ is the chemical potential, $\Delta_{t,b}$ are proximity induced pairing gap functions on $t$ and $b$ surfaces, which are chosen as $\vec{k}$ independent due to $s$-wave superconducting proximity effect. We consider low-temperature cases, and when the in-plane field $|\vec{H}|$ is smaller than the upper critical field of the parent superconductor, $\Delta_i$ remains finite and does not significantly change. The possible interlayer pairing is studied in Supplemental Materials~\cite{supple}.

\begin{figure}[t]
\begin{center}
\includegraphics[width=2.2in,clip=true]{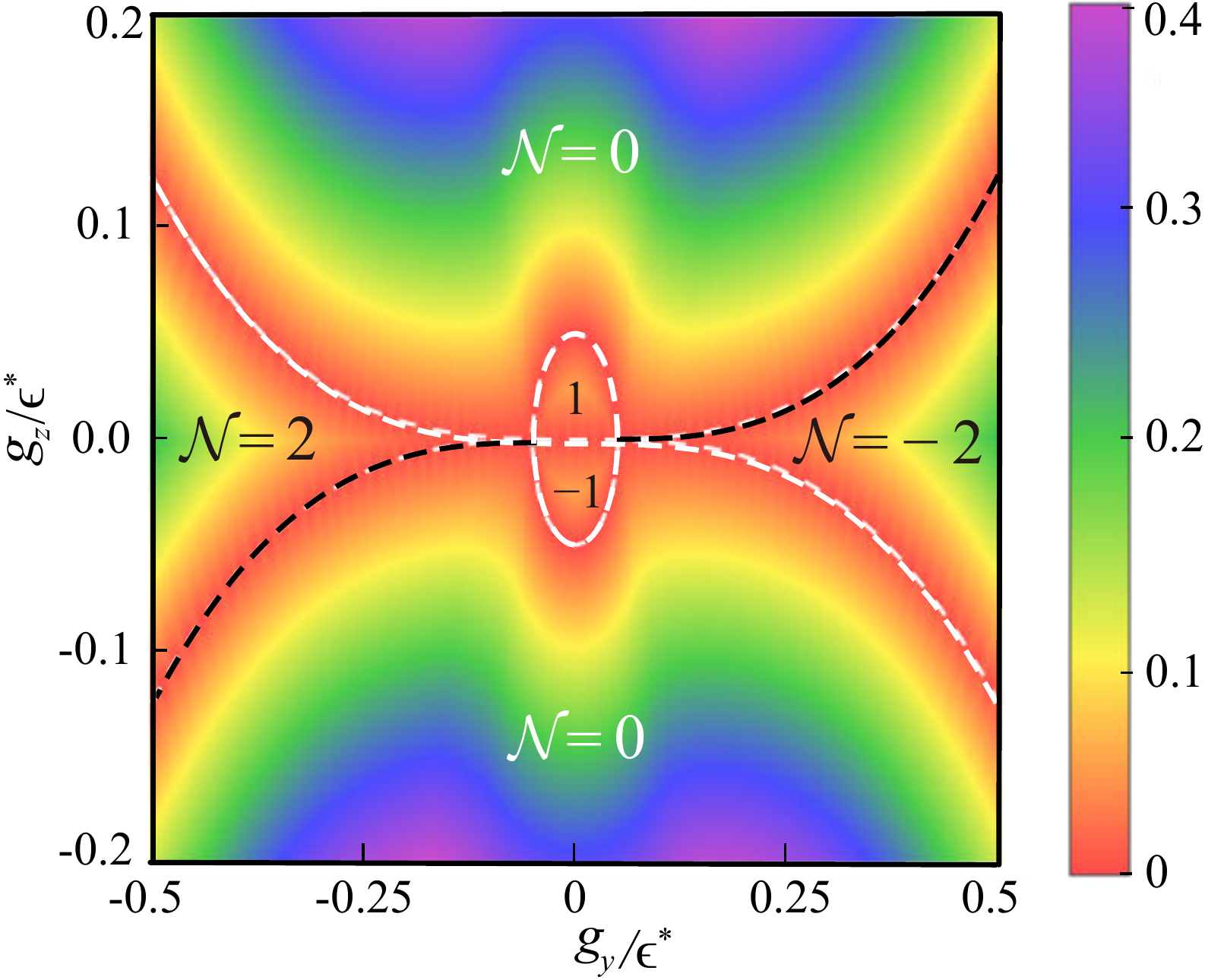}
\end{center}
\caption{Phase diagram of even SL MnBi$_2$Te$_4$-superconductor hybrid system for $\mu=0$, $\Delta_b=0$, and $\Delta_t/\epsilon^*=0.05$. The colour represents the BdG gap. Here $\Delta_t$ is chosen to be un-physically large to see the small odd $\mathcal{N}$ TSC regions.}
\label{fig4}
\end{figure}

The optimal condition for realizing the $\mathcal{N}=\pm1$ TSC is to have inequivalent pairing on the two surfaces~\cite{wang2015c}. Thus here we only plot the phase diagram for $\mu=0$ and $\Delta_b=0$ in Fig.~\ref{fig4}. One can see that only within the small circle around $\vec{g}=0$ defined by $|\Delta_t|=\sum_i\sqrt{g_{i}^2}\equiv g$, the $\mathcal{N}=\pm1$ TSC is realized. The phase boundary between $\mathcal{N}=0$ and $\mathcal{N}=\pm2$ in other regions is roughly the same as boundary between $C=0$ and $C=\pm1$ in Fig.~\ref{fig1}(c). This is simply because the in-plane field $\vec{g}_{\parallel}$ shift the entire Fermi surface in the perpendicular direction in the Brillouin zone, and the energy between states at $\vec{k}$ and $-\vec{k}$ no longer degenerate and lead to pair breaking effect. When $g>|\Delta_t|$, the transitions are degenerate, namely directly from $\mathcal{N}=\pm2$ to $\mathcal{N}=0$ without intermediate phase. Quite different from Ref.~\cite{wang2015c}, where finite $\mu$ will enlarge $\mathcal{N}=\pm1$ phases. Here finite $\mu$ will lead to metallic state in bottom surface, and the top surface enters into a gapless superconductor with partial Bogoliubov Fermi surface~\cite{yuan2018,supple}. From the example studied above, we conclude that chiral TSC does not necessarily emerge at QAH plateau transition if the exchange field is strong.

\emph{Discussion.} Finally we discuss the experimental feasibility. (i) Structure inversion asymmetry $\delta V$ between the two surfaces should be smaller than $\text{max}(m_t',m_b')$, then the field-induced QAH transition suivives. (ii) We estimate $H^*_{\parallel}$ and the QAH gap. Obviously, $H^*_{\parallel}$ depends on the field direction relative to crystalline orientation from Fig.~\ref{fig1}(b). Take $\Gamma M$ for example, the transition is at $|g_z|=|g_y^3/\epsilon^{*2}|$. By assuming $J_z=J_{\parallel}$, then $\cos\phi^*=0.95$ determines $H^*_{\parallel}$. $\phi$ is the angle between magnetic moment and $H$. $M_{\parallel}$ is linear in $H$, i.e., $S_{\text{sat}}\cos\phi\propto H$, where $S_{\text{sat}}\approx3.6$ is the saturation magnetic moment~\cite{yan2019b}. Thus, $H^*_{\parallel}\approx8.6$~T obtained when in-plane moment roughly equals $S_{\text{sat}}\cos\phi^*$~\cite{otrokov2018,supple}. The estimated QAH gap is $2g_z^3/\epsilon^{*2}\approx2.8$~meV~$\approx 33$~K. The large surface gap in MnBi$_2$Te$_4$ $g_z\approx50$~meV~\cite{vidal2019} makes the in-plane QAH transition feasible in experiment, which is impossible for dilute magnetic TIs. (iii) The above study can be directly applied to other magnetic TI system such as MnBi$_4$Te$_7$ and MnBi$_6$Te$_{10}$~\cite{aliev2019,hu2019,wu2019}. The AFM coupling and uniaxial anisotropy in these two materials are weaker compared to MnBi$_2$Te$_4$, which leads to a smaller critical $H^*_{\parallel}$~\cite{supple}.  (iv) The out-of-plane field-induced QAH plateau transition found in Ref.~\cite{liu2019} is similar to the case studied here, where AFM multidomains \emph{spin-flop} and cant. At the spin-flop field, the system is described by Eq.~(\ref{single}) but with random $\vec{g}_{\parallel}$ and $g_z\approx0$. By further increasing the field, $\vec{g}$ cants along $z$-axis and induces $0$ to $\pm1$ transition. The estimated critical field $H^*_{\perp}\approx4.5$~T~\cite{supple}, which is consistent with experimental value $4.58$~T~\cite{liu2019}. It is worth mentioning that the plateau transition in FM TIs always accompany the coercivity transition, where the abrupt coercivity transition may completely conceal the universal scaling of the QAH plateau transition~\cite{wang2014a}. Here there is no coercivity transition in MnBi$_2$Te$_4$ due to AFM ordering. (v) Lastly, MnBi$_2$Te$_4$ may not be suitable for chiral TSC due to strong exchange field. However, its descendent systems MnBi$_4$Te$_7$ and MnBi$_6$Te$_{10}$ thin films may be good platforms for chiral Majorana fermion. There, one surface of Bi$_2$Te$_3$ is gapped by proximity coupled to superconductor, and the other surface is magnetically gapped by MnBi$_2$Te$_4$.

\begin{acknowledgments}
J.W. acknowledge Jiang Xiao for helpful discussions. This work is supported by the Natural Science Foundation of China through Grant No.~11774065, the National Key Research Program of China under Grant No.~2016YFA0300703, the Natural Science Foundation of Shanghai under Grant No.~17ZR1442500, 19ZR1471400. J.Z. and Z.L. contributed equally to this work.
\end{acknowledgments}

\end{document}